\documentclass[fleqn,usenatbib]{mnras}
\pdfoutput=1
\usepackage{newtxtext,newtxmath}

\usepackage[T1]{fontenc}

\DeclareRobustCommand{\VAN}[3]{#2}
\let\VANthebibliography\thebibliography
\def\thebibliography{\DeclareRobustCommand{\VAN}[3]{##3}\VANthebibliography}

\usepackage{graphicx}	
\usepackage{amsmath}	
\usepackage{booktabs}
\usepackage{placeins}
\usepackage{multirow}
\usepackage{subfig}
\usepackage{caption,booktabs}

\title[Signal Demodulation with a CRHWP]{CMB Polarisation Signal Demodulation with a Rotating Half-Wave Plate}

\author[M. Rashid, M. L. Brown, D. B. Thomas]{
Mariam Rashid$^{1}$,  Michael L. Brown$^{1}$, Daniel B. Thomas$^{1, 2}$
\\
$^{1}$Jodrell Bank Centre for Astrophysics, Department of Physics and Astronomy, School of Natural Sciences, The University of Manchester, \\ Oxford Road, Manchester M13 9PL, UK\\
$^{2}$ Department of Physics and Astronomy, Queen Mary University of London, UK. 
}

\date{Accepted XXX. Received YYY; in original form ZZZ}

\pubyear{2023}

\begin{document}
\label{firstpage}
\pagerange{\pageref{firstpage}--\pageref{lastpage}}
\maketitle

\begin{abstract}
Several forthcoming Cosmic Microwave Background polarisation experiments will employ a Continuously Rotating Half-Wave Plate (CRHWP), the primary purpose of which is to mitigate instrumental systematic effects. The use of a CRHWP necessitates demodulating the time-ordered data during the early stages of data processing. The standard approach is to ``lock in'' on the polarisation signal using the known polarisation modulation frequency and use Fourier techniques to filter out the remaining unwanted components. However, an alternative, less well-studied option is to incorporate the demodulation directly into the map-making step. Using simulations, we compare the performance of these two approaches to determine which is most effective for $B$-mode signal recovery. Testing the two techniques in multiple experimental scenarios, we find that the lock-in technique performs best over the full multipole range explored. However, for the recovery of the largest angular scales (multipoles, $\ell < 100$) we find essentially no difference in the recovery of the signal between the lock-in and map-making approaches, suggesting that a parallel analysis based on the latter approach could represent a powerful consistency check for primordial $B$-mode experiments employing a CRHWP. We also investigate the impact of a detector-differencing step, implemented prior to demodulation, finding that, in most scenarios, it makes no difference whether differencing is used or not. However, analysing detectors individually allows the point at which information from multiple detectors is combined to be moved to later stages in the analysis pipeline. This presents alternative options for dealing with additional instrumental systematic effects that are not mitigated by the CRHWP.  

\end{abstract}

\begin{keywords}
Cosmology -- Cosmic Microwave Background -- Polarisation --- Data Analysis 
\end{keywords}



\section{Introduction}


%

The polarisation of the Cosmic Microwave Background (CMB) provides unique and valuable information about the early universe \citep[see e.g.][]{polarisation_primer}. In particular, many inflation theories predict a potentially detectable $B$-mode polarisation signal in the CMB, imprinted by primordial gravitational waves at early times. Thus, detection of this $B$-mode signal could provide strong evidence in favour of inflation \citep{grav_waves_primer, bmode_challenges}. However there are several challenges associated with its detection. Most notably, the signal is very small and could easily be obfuscated by incorrectly subtracted foregrounds, inaccurately modeled detector noise, instrumental polarisation (temperature to polarisation leakage) or $B$-modes generated by gravitational lensing of the CMB $E$-mode signal \citep{Essinger_Hileman_2016, Ade_2014, shimon08, bmode_challenges, LEWIS_2006, planck_2016_XLI, 2014PTEP.2014fB109I, prunet1999polarized, Weiland_2022, Simon_2018, McCallum_2020, hu03, odea07}. 

One way of overcoming many of these challenges is to use a continuously rotating half-wave plate (CRHWP), as previously implemented in experiments such as MAXIPOL \citep{maxipol_hwp}, ABS \citep{ABS_1} and POLARBEAR \citep{polarbear_1}. Several forthcoming experiments, most notably Simons Observatory \citep[SO][]{SO_Main} and LiteBIRD \citep{LiteBIRD_1}, will also use a CRHWP. 

A CRHWP offers three major benefits. One advantage of a rotating waveplate is that it increases the variety of polarisation angles measured by a detector as it passes over a pixel. Secondly, it reduces the impact of low-frequency noise by modulating the signals of interest into a higher frequency band. Finally, it allows a detector that is sensitive to polarisation in a single direction to measure the polarisation in multiple directions as it rotates \citep{CRHWP_polarbear}. In principle, this allows the total intensity (or temperature, $T$) and two linear polarisation Stokes parameters ($Q$ and $U$) to be measured using a single detector, which could be beneficial for mitigating the impact of temperature-to-polarisation ($T \rightarrow P$) ``leakage'' effects associated with mismatches in the responses of different detectors \citep[e.g.][]{Essinger_Hileman_2016}.
However, a number of issues and open questions surrounding the use of CRHWPs in CMB experiments remain \citep{SAT_Telsecopes_1,Essinger_Hileman_2016,CRHWP_polarbear,ACT_HWP,SO_HWP_systematics}. The question this paper aims to address is that of how best to demodulate the signal.

Demodulation is the process of recovering the targeted temperature and polarisation signals from a timestream that has been subject to rapid modulation (e.g through the use of a CRHWP). The first CMB experiment to implement rapid modulation with a CRHWP was MAXIPOL, the analysis of which used a classical ``lock-in'' technique, followed by the application of a low-pass filter in Fourier space \citep{maxipol_hwp}. This approach has become the standard way to demodulate CMB experiments employing a CRHWP with the more recent ABS and POLARBEAR experiments using similar techniques \citep{ABS_1, polarbear_demod}. Here we compare the efficacy of this standard method to a demodulation method implemented directly in the time domain. We also examine the effect of detector differencing on the two demodulation methods: differencing pairs of detectors that are oriented $90^\circ$ apart is commonly used in the absence of a HWP as (in the absence of systematics) doing so removes the temperature signal as described in section \ref{sec:differencing}. Such differencing has been used very successsfully in South Pole experiments that have not included a HWP \citep[e.g. the QUaD and BICEP/Keck series of experiments,][]{2009ApJ...692.1247P, 2009ApJ...705..978B, 2010ApJ...711.1123C, 2014PhRvL.112x1101B}. However, the advantages of detector differencing is less well studied for experiments that include a HWP.

In this work, we demonstrate our techniques using simulations, for which we assume a SO-like setup. SO is under construction in the Atacama Desert of Chile, and aims to constrain the tensor-to-scalar ratio $r$ to a precision of at least $\sigma(r) = 0.003$ \citep{SO_Main}. To this end, SO plans to deploy three Small Aperture Telescopes (SATs) by mid-2024 \citep{SO_Main}, to be expanded to six SATs by mid-2026. All of the SO SATs will employ cryogenic CRHWPs, and are designed to probe a multipole range of $30 < \ell < 300$.


In section \ref{sec:demods}, we introduce the demodulation procedures investigated in this paper. We then describe how the simulations were generated in section \ref{sec:sims}. In section \ref{sec:analysis}, we describe our data analysis steps and in section \ref{sec:results}, we discuss the relative success of each pipeline in the six different noise regimes. In section \ref{sec:diff_effect}, we explore the effect of combining data from two detectors which have not been differenced. Finally, section \ref{sec:conclusion} contains our conclusions. 

\section{Demodulation Procedures}
\label{sec:demods}

Two demodulation procedures are investigated in this work -- a technique that works directly on the time-domain data in real space, which we refer to as Time Domain Demodulation (TDD, section \ref{sec:timedemod}) and the classical lock-in technique employing a low-pass filter applied in frequency space, which we refer to as Frequency Domain Demodulation (FDD, section \ref{sec:freqdemod}). Both approaches are tested, with and without prior detector differencing, as described in section \ref{sec:differencing}. 

The (noiseless) time-ordered data (TOD) signal seen by a single detector can be expressed as
\begin{equation}
    d_i = \frac{1}{2} \left[T + Q'\cos(4 \phi_i) + U'\sin(4 \phi_i)\right],
    \label{data eqn0}
\end{equation}
where $\phi$ is the combination $\beta + \theta / 2$ of the half-wave plate angle $\beta$ and the angle of the detector on the sky $\theta$. The Stokes parameters, $T, Q'$ and $U'$, describe the on-sky signal containing only the CMB signal. No foregrounds are considered in this work, and as such all references to 'signal-only' simulations henceforth refer to the CMB-only case. In the absence of foregrounds it is also sufficient to take a single frequency. The 93GHz band of the SO SATs is modelled throughout.  The $i$ index denotes the individual data samples in the timestream. If one considers the data collected over a sufficiently short duration such that the angle that the detector makes with respect to the sky is approximately constant (e.g. a single constant elevation scan for a typical ground-based CMB experiment), then equation~(\ref{data eqn0}) can be written as 
\begin{equation}
    d_i = \frac{1}{2} \left[T + Q\cos(4 \beta_i) + U\sin(4 \beta_i)\right],
    \label{data eqn}
\end{equation}
where the polarisation Stokes parameters $Q$ and $U$ are now defined with respect to a reference coordinate axis aligned with $\theta$. In what follows, we assume that all demodulation procedures are  applied to such short segments of data (so that variations of the detector angle on the sky are not considered). In addition, the simulations we use to demonstrate our techniques do not incorporate any variation in detector angles with respect to the sky. We also do not consider any non-idealities in the half-wave plate, and therefore assume perfect behaviour of the rotation angle $\beta$. In the case where there are significant variations of the detector angle over the duration of a scan (which is a possibility for realistic long-duration scans), then these variations can be readily incorporated, either as part of the demodulation procedure itself, or as part of the projection of the TOD into maps, subsequent to demodulation. 

\subsection{Time Domain Demodulation}
\label{sec:timedemod}

The TDD method averages consecutive data samples across a user-defined timescale labelled $\tau_{\rm dm}$, where the subscript indicates this is a \textit{demodulation} parameter. We make the assumption that the on-sky signal under investigation ($T$, $Q$ or $U$) does not change significantly over this time, with $\tau_{\rm dm}$ sufficiently short to justify this. As $\tau_{\rm dm}$ must include at least three data samples in order to fit for the three Stokes parameters in the single-frequency case, $\tau_{\rm dm} > 3 / f_{\rm samp}$. Estimates of $T, Q$ and $U$ are found using a `best fit' approach, wherein for each time segment we estimate the values of $T, Q$ and $U$ as linear combinations of weighted averages of the data over that time. To derive this estimator, one begins by defining a $\chi^2$ value: 
\begin{equation}
    \chi^2 = \sum_i w_i \left(d_i - \frac{1}{2} \left[T + Q\cos(4 \beta_i) + U\sin(4 \beta_i)\right]\right)^2, 
\end{equation}
where $w_i$ is an arbitrary weight, requiring only a normalisation such that $\sum_i w_i =1$. Minimising this with respect to the Stokes parameters yields the following expressions for $T, Q$ and $U$:
\begin{equation}
\begin{aligned}
    \begin{pmatrix}T \\ Q \\ U \end{pmatrix} = 
    {} & 2 \begin{pmatrix} 1 & \langle \cos(4\beta_i) \rangle & \langle \sin(4\beta_i) \rangle \\
    \langle \cos(4\beta_i) \rangle & \langle \cos^2(4\beta_i) \rangle & \langle \cos(4\beta_i)\sin(4\beta_i) \rangle \\
    \langle \sin(4\beta_i) \rangle & \langle \cos(4\beta_i)\sin(4\beta_i) \rangle & \langle \sin^2(4\beta_i) \rangle \end{pmatrix}^{-1} \\
    & \cdot \begin{pmatrix} \langle d_i \rangle \\ \langle \cos(4\beta_i)d_i \rangle \\ \langle \sin(4\beta_i) d_i\rangle \end{pmatrix}.
\end{aligned}{}
\label{eq:tdd}
\end{equation}
Here, $\langle \cdots \rangle$ denotes the average over the (weighted) data samples within a time period $\tau_{\rm  dm}$. 

We have determined the value of $\tau_{\rm dm} = 5/ f_{\rm fsamp}$ as the optimal choice for this parameter value. This value minimizes the impact of a $E\rightarrow B$ leakage effect that is present in the TDD analysis (see Section\ref{sec:signOnly}). We therefore set $\tau_{\rm dm} = 5 /f_{\rm samp}$ for all of our analysis.

One can estimate the scale at which the assumption inherent in the TDD approach is likely to break down for this choice (i.e. the scale below which the assumption of constancy of the on-sky signals is no longer valid). For the telescope scan speed and sampling frequency used in our simulations (see Table~\ref{tab:params}), the angular scale corresponding to $\tau_{\rm dm}$ is $\sim\!3.75$ arcmin, corresponding to multipoles, $\ell > 2800$, well beyond the range of scales included in our simulations, or those of interest for $B$-mode measurements. 

Demodulation in the map-making step could be implemented differently to the TDD approach used here, in that all the hits within a given pixel could be included in a single pointing matrix for that pixel, and the T/Q/U solution in a pixel created in a single step (rather than the two-step process in the TDD method we've laid out here). One advantage of the two step process is that for rapid modulation the angular scale over which the sky signal is assumed to be constant is much smaller. We leave to future work a detailed investigation of whether the two-step process could be related to the leakage seen in the TDD method (see Section \ref{sec:results}).

\subsection{Frequency Domain Demodulation}
\label{sec:freqdemod}
The FDD method uses reference signals to lock in to the signal of interest (either $T$, $Q$ or $U$) contained within the modulated timestream. The targeted signal can then be extracted using a low pass filter in frequency space. For example, to lock in to the Stokes $Q$ polarisation signal, one multiplies the modulated TOD by a reference signal of $2\cos(4\beta_i)$. Multiplying equation~(\ref{data eqn}) by this term yields
\begin{equation}
    2\cos(4\beta_i)d_i  = \frac{1}{2} \left[Q + Q\cos(8\beta_i) + 2T\cos(4\beta_i) + U\sin(8\beta_i) \right],
    \label{Qmod}
\end{equation}
resulting in a TOD for which the unmodulated signal component is entirely Stokes $Q$. Similarly to recover the $U$ polarisation, the TOD is multiplied by a reference signal of $2\sin(4\beta_i)$:
\begin{equation}
        2\sin(4\beta_i)d_i = \frac{1}{2} \left[U - U\cos(8\beta_i) + 2T\cos(4\beta_i) + Q\sin(8\beta_i) \right].
    \label{Umod}
\end{equation}
%
%
Finally to remove the unwanted oscillatory terms from the locked-in signals a low pass filter is applied i.e. the data is transformed to Fourier space where all modes with frequency larger than a user-defined cut-off frequency are set to zero.  In this paper we have used a sharp cut-off in frequency space at $f_{\rm cutoff} = 3 f_\lambda$, where $f_\lambda$ is the HWP rotation frequency. We note that other choices of filter could potentially improve the covariance properties of the resulting demodulated, low-pass filtered data. 


\subsection{Detector Differencing}
\label{sec:differencing}
Differencing the TODs from orthogonal pairs of co-located detectors is a common technique used to reject the total intensity ($T$) signal when polarisation is the primary signal of interest. If there are two detectors (A and B) oriented at $90^\circ$ to each other (i.e. the detector angle for detector B is given by $\theta^B_i=\theta^A_i+90^\circ$), and the half wave plate angle $\beta$ is the same for both detectors, then, using equation~(\ref{data eqn}), the differenced signal ($d^{\rm A-B} = d^{\rm A} - d^{\rm B}$) is 
\begin{equation}
    d^{\rm A-B}_i = Q\cos(4\beta_i) + U\sin(4\beta_i),
\end{equation}
isolating the polarisation. In practice, if the detector response functions are different (e.g. differential gain, differential pointing or differences in the beam responses of the two detectors), this can cause temperature-to-polarisation leakage, biasing the polarisation recovery \citep{odea07, hu03, shimon08, Wallis_2016}. As part of our analysis, we investigate whether or not differencing detectors in advance of applying the demodulation step is beneficial to the recovery of the polarisation signal.

\section{Simulations}
\label{sec:sims}
To investigate the different approaches described above, we simulate the time-ordered-data for a single pair of detectors which are sensitive to orthogonal polarisation directions (nominally aligned with the North-South and East-West axes of the input \texttt{HEALPIX} maps. The details of the simulations are as follows.

\subsection{Input CMB maps}
We generate model $TT$, $EE$, $TE$ and $BB$ CMB power spectra using \texttt{CAMB} \citep{lewis00}. The input cosmology used consisted of the best-fit standard $\Lambda$CDM model to the 5-year \emph{WMAP} data set \citep{hinshaw09}, with the following values for the cosmological parameters: $\Omega_b = 0.044, \Omega_{\rm cdm} = 0.212, \Omega_\Lambda = 0.744, H_0 = 72 \, \mbox{km s}^{-1} \, \mbox{Mpc}^{-1}, T_{\rm CMB} = 2.726 \, K, n_s = 0.963, \tau = 0.087$. The input tensor-to-scalar ratio was set to $r=0.026$. Gaussian realisations of $T$, $Q$ and $U$ maps were then generated from these power spectra using the \texttt{SYNFAST} facility of the \texttt{HEALPIX}\footnote{See \url{http://healpix.sourceforge.io}} software \citep{healpix}. The input $BB$ power spectrum includes the expected $B$ mode signal arising from weak gravitational lensing effects (i.e. we model the lensing $B$-modes as a purely Gaussian field). As part of the simulation process, the input CMB signal is convolved with a perfect Gaussian beam with FWHM $=30$ arcmin, matching the beam size of the SO SATs at 93 GHz \citep{SAT_Telsecopes_1}. These input maps were created using a  \texttt{HEALPIX} \texttt{NSIDE} parameter of 2048, which corresponds to a mean map pixel scale of $1.7$ arcmin \citep{healpix}.

\subsection{Simulated detector data -- signal component}
Simulated TODs are created by scanning across and sampling from the input \texttt{HEALPIX}-generated maps described above. Key parameters describing our TOD simulations are listed in Table \ref{tab:params}. 
\begin{table}
\centering
\caption{Parameter values chosen for the simulations in this paper. Note that the size of the differential pointing error, and the polarisation fraction $f_{\rm pol}$, are both unrealistically large and have been used for illustrative purposes only.}
\begin{tabular}{|l|l|l}
\cline{1-3}
\textbf{Parameter}              & \textbf{Value}     &
\\ \cline{1-3}
Relative gain errors            & 1\%                &                     
\\ 
White noise RMS ($\sigma_{\rm rms}$)                 & 2.25 $\mu$K-arcmin &                                                                                          \\ 
$1/f$ noise knee frequency ($f_{\rm knee}$)      & 0.45 Hz            & 
\\ 
$1/f$ noise power law index ($\alpha$)     & 2.5                &                                                                                          \\ 
Polarization fraction (for polarised $1/f$ noise, $f_{\rm pol}$)     & 0.5               &                                                                                          \\ 
HWP rotation frequency ($f_\lambda$)          & 3 Hz               & 
\\ 
Sampling frequency ($f_{\rm samp}$)              & 60 Hz              & 
\\ 
Telescope scan speed            & 0.75 deg/sec       &   

\\ 
Differential pointing error (offest in both R.A. \& Dec.)           & 1.7 arcmin       &  

\\ \cline{1-3}
\end{tabular}
\label{tab:params}
\end{table}
%
The scan covers a roughly square area of the sky, 66 degrees on a side. With the apodisation applied (to reduce edge effects; see section \ref{sec:analysis}), the effective map area is roughly 65 $\times$ 65 deg$^2$. Approximately 10\% of the sky is therefore retained, in broad agreement with the expected survey area for the SO deep survey \citep{SO_Main}. The targeted sky region is scanned in rows of constant declination, simulating a detector which begins observations at the Eastern edge of the survey region and pans across at a constant scanning speed to the Western edge of the survey area. The detector then repeats the exact same pattern for the next step in declination. A declination step size of 2.28 arcmin is used. The scan speed and sampling rate are both kept constant for the entire scan resulting in fewer data points closer to the poles. The scan strategy modelled here is unrealistic in practice for a ground-based observatory located in Chile using constant elevation scans. In particular, the effects of sky rotation are not accounted for. However, this model is sufficient for our purposes as the limited improvement in crossing angle coverage achievable for constant-elevation scans from an observing site such as the SO site in Chile would result in only a very marginal improvement in the polarisation angle coverage of an experiment employing a CRHWP \citep{2021OJAp....4E..10T}. In other words, for a given detector pair and a scan strategy based on Constant Elevation Scans, the crossing angle variation from sky rotation will be small, and the CRHWP will dominate the spread of angles in equation \ref{eq:tdd} for any given pixel. In addition, we are not using any techniques (such as destriping) where the relative directions of the scanning tracks on the sky are important for the method. For these reasons, we believe that the idealised scan strategy used in this paper is sufficient to elucidate the key differences between the demodulation techniques that we investigate here.

Simulated signal-only (i.e. CMB-only) TODs are generated using equation~(\ref{data eqn}). 
The HWP angle and pointing information for each data point is calculated using the scan speed, sampling rate and HWP rotation frequency. The $T$, $Q$ and $U$ signal values for each TOD value are inferred from the input sky maps using bi-linear interpolation via the \texttt{HEALPY} \texttt{get\_interp\_val} method. The HWP is taken to be ideal, with no irregularity; the effects of irregularity and their interaction with the demodulation techniques studied here will be investigated elsewhere (Rashid, in prep.).

\subsection{Including noise}
\label{sec:noise}
We test the performance of the demodulation pipelines in the presence of three different types of noise: uncorrelated Gaussian-distributed white noise, unpolarised $1/f$ noise and polarised $1/f$ noise. The $1/f$ unpolarised case is included to approximate the effect of common-mode fluctuations in the atmosphere and we include the polarised $1/f$ case to explore the sensitivity of our pipelines to a polarised component in the atmospheric fluctuations. In the presence of all three of these noise sources, the TOD for each detector within our pair is generated as 
\begin{multline}
d_i = \frac{1}{2} \left\{T + \left[Q + n^Q_i\right] \cos(4\beta_i) + \left[U + n^U_i\right] \sin(4\beta_i) \right\} \\ 
+ n^{\rm white}_{i} + n^{\rm non-white}_i,
\label{data_noisy}
\end{multline}
where $n^{\rm white}_i$ is the white noise, $n^{\rm non-white}_i$ is the unpolarised $1/f$ noise and $\{n^Q_i, n^U_i\}$ are the polarised $1/f$ noise.

The white noise, $n^{\rm white}_i$, is generated as zero-mean random Gaussian deviates with a standard deviation ($\sigma_{\rm rms}$) tuned to ensure a white noise level in the resulting maps of 2.25 $\mu$K-arcmin. This final map-noise level is chosen to match the expected performance of the SO SATs assuming a five year survey \citep{SO_Main}. The white noise is taken to be fully uncorrelated between the two detectors, i.e. two independent sets of random Gaussian deviates (with the same standard deviation) are generated, one for each detector in the pair. 

The atmospheric unpolarised $1/f$ noise, $n_i^{\rm non-white}$, is also modeled as Gaussian-distributed, but with a power spectrum,
\begin{equation}
        P(f) = \sigma_{\rm rms}^2 \left(\frac{f_{\rm knee}}{f}\right) ^\alpha,
\end{equation}
where $\sigma_{\rm rms}$ is the standard deviation of the Gaussian white noise, and $f_{\rm knee}$ and $\alpha$ are the knee frequency and spectral index of the $1/f$ noise respectively. Following \cite{HWPOverheat_Michael}, we set $f_{\rm knee} = 0.45$ and $\alpha = 2.5$ which provides a good fit to the atmospheric $1/f$ noise seen in the QUaD data \citep{2009ApJ...705..978B}. The $1/f$ noise is assumed to be fully correlated between the two detectors (which we take to be co-located). It is therefore generated only once, and is added to both detector TODs. 


Polarised $1/f$ noise, $\{n^Q_i, n^U_i\}$, is generated in much the same way, but with a power spectrum 
\begin{equation}
        P_{\rm pol}(f)= f_{\rm pol} P(f)
\end{equation}
where $f_{\rm pol}$ is the assumed polarisation fraction of the atmospheric noise, taken to be 0.5. To simulate a polarised component, two arrays of random numbers are required: $n_i^Q$ to be added to the on-sky Q signal and $n_i^U$ which is added to the U signal (see equation \ref{data_noisy}). As with the unpolarised $1/f$, this polarised component is taken to be fully correlated between detectors so the same two sets of random numbers ($n_i^Q$ and $n_i^U$ ) are added to both detectors. Note however that, unlike the common-mode $1/f$ noise and the uncorrelated white noise, the polarised $1/f$ noise is modulated by the CRHWP. 

\subsection{Instrumental systematics}
\label{sec:systematics}
It is also interesting to explore which (if any) of the considered demodulation pipelines performs best in the presence of instrumental systematic errors. Previous studies have explored in detail the degree to which fast modulation mitigates instrumental systematic effects \citep{2008ApJ...689..655M, HWPOverheat_Michael}. Our goal here is not to replicate such studies but to identify a preferred approach to performing the demodulation step in the presence of real-world effects. We therefore restrict the current study to explore example systematics for which one may expect to see a difference between, for example, a demodulation approach based on differencing and one based on analysing individual detectors in isolation. We investigate two such effects in this paper -- a differential gain error between the two detectors within a pair and a differential pointing error. 

To model a differential gain error, the response of one of the two detectors is amplified by 1\%. To model a differential pointing error between the detectors, for both detectors within a pair, we apply an offset of 1.7 arcminutes in both latitude and longitude directions. 


\section{Data analysis}
\label{sec:analysis}
The TOD generated by each simulation is processed through both TDD and FDD pipelines (see Section~\ref{sec:demods}). To investigate the impact of detector differencing, in addition to processing the individual simulated TODs for the two detectors, we also subtract these to produce a differenced TOD.  This is then propagated through the demodulation pipelines alongside the analysis of the single-detector TODs. 

The implementation of the TDD for the differenced case proceeds in an identical fashion to equation~(\ref{eq:tdd}). Note in particular that, following \cite{2014ApJ...794..171P} and  \cite{mccallum21}, we retain the $T$ elements of this matrix application to the (weighted) data in order to solve for any $T \rightarrow P$ leakage that may be present in the differenced data (e.g due to gain errors). The FDD analysis for the differenced data also proceeds as in the undifferenced case (see equations~\ref{Qmod} and \ref{Umod}). Obviously, for a perfectly calibrated detector pair, with identical beam response functions, the modulated $T$ terms on the RHS of these equations would not be present. 

In most cases where we demodulate the two detectors independently, we have also included a final averaging step -- where we have taken the average of the pairs of $T$, $Q$ and $U$ TODs resulting from the demodulation of the two detectors. In addition, for a restricted set of simulations, we will also present results arising from the analysis of single detectors in isolation (without any subsequent averaging).

Once demodulated (and, where relevant, averaged), the data is projected into maps of $T$, $Q$ and $U$ using simple binning (where each pixel value is determined as the average of all time-ordered data points falling within that pixel). We note that simple binning is not optimal in the presence of correlated noise, and map-making techniques to minimise the effect of correlated noise have been investigated in other work \citep[see e.g.][]{10.1111/j.1365-2966.2010.16954.x}. We use a standard pseudo-$C_\ell$ analysis \citep{pseudo_Cl} to extract the $E-$ and $B-$mode power spectra from the resultant maps, using the \texttt{HEALPIX} \emph{anafast} utility. For the power spectrum extraction, a uniform mask (weight function) is adopted, the boundaries of which were softened with a simple cosine apodisation of width $\sim$1 deg. 

For simulations including noise, in order to remove the noise bias, the power spectra are extracted using a cross-correlation approach. Two independent realisations of the noise are generated following the procedures described in Section~\ref{sec:noise}. These are then added to a single realisation of the signal to produce two sets of signal + noise TODs (and subsequently two sets of $T$, $Q$ and $U$ maps) which contain an identical  signal component but different realisations of the noise (though based on the same noise model). The power spectra, calculated as the cross-correlation of these two sets of maps, will then be insensitive to the noise bias. 

\section{Results}
\label{sec:results}
For each noise model and/or systematic case investigated, we generate and analyse 25 realisations. This results in four sets of 25 recovered $E$- and $B$-mode power spectrum estimates, corresponding to our four demodulation pipelines (undifferenced TDD, differenced TDD, undifferenced FDD, differenced FDD). In each case, we take the average of the 25 recovered power spectra, with error bars estimated as the standard deviation of the 25 runs. The data is binned with bin widths of $\Delta\ell = 10$, and plotted at the central value of the bin.

In order to quantify the success of the demodulation pipelines, we compare the recovered $B$-mode pseudo-$C_\ell$ to the mean pseudo-power spectra measured from 25 signal-only map-based simulations (i.e. where no TOD generation or demodulation procedure has been applied). We refer to these simulations hereafter as "simple simulations". These simple simulations consisted simply of the generation of $T$, $Q$ and $U$ maps with the \texttt{HEALPIX} \texttt{SYNFAST} facility, from which the pseudo-power spectra were then directly measured with \texttt{ANAFAST} (adopting the same mask as was used for the analysis of the full TOD simulations). 

As a preliminary check that the pseudo-power spectra extraction is behaving as expected, we can compare the mean spectrum recovered from our  simple simulations with the predicted theoretical signal, taking into account the effect of the sky mask. This predicted signal is calculated as 
\begin{equation}
\widetilde{C}^{BB}_{\ell} = \sum_{\ell'} \left\{ M^{EE \rightarrow BB}_{\ell\ell'} C^{EE}_{\ell'} + M^{BB \rightarrow BB}_{\ell\ell'} C^{BB}_{\ell'} \right\},
\label{eq:pcl_theory}
\end{equation}
where $M^{EE \rightarrow BB}_{\ell\ell'}$ and $M^{BB \rightarrow BB}_{\ell\ell'}$ are the relevant parts of the pseudo-$C_\ell$ coupling matrix, which is, in turn, calculable from the mask \citep{pseudo_Cl}. 

\begin{figure}
    \centering
    \includegraphics[width=0.45\textwidth]{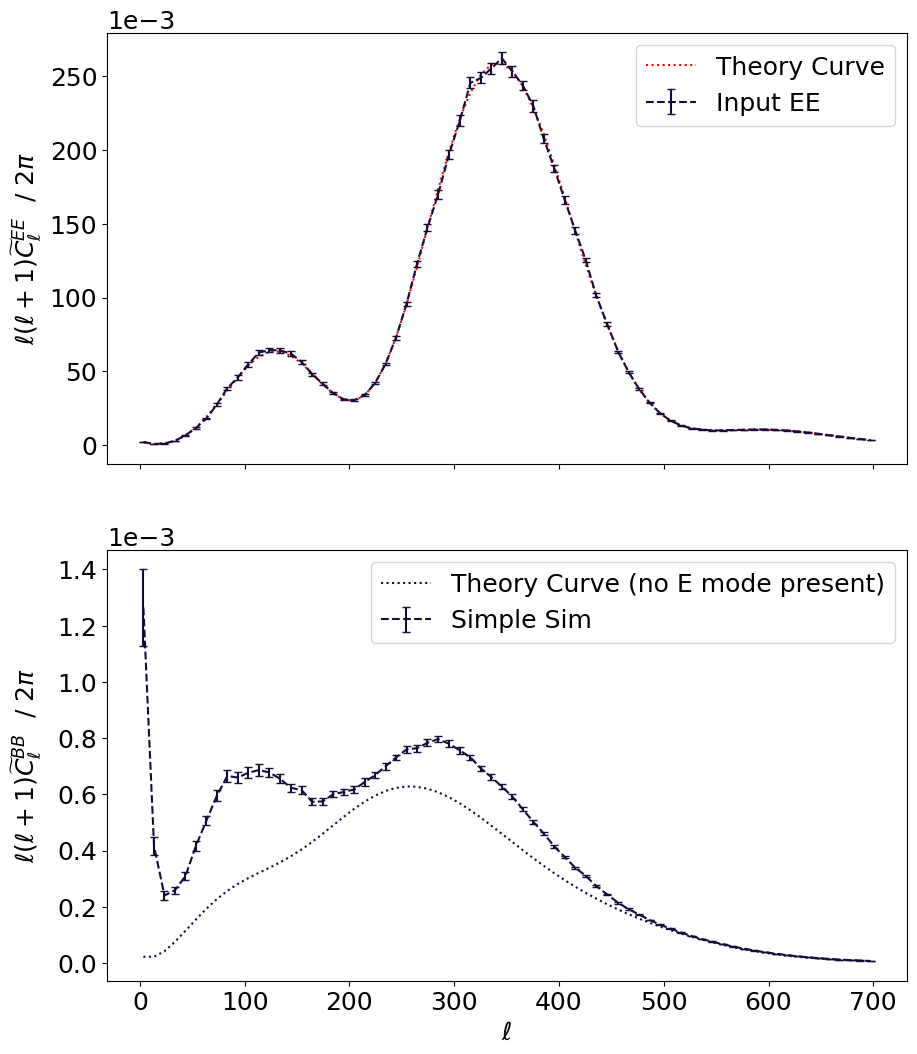}
    \caption[Comparison of the predicted pseudo-power spectra (dotted red curves) with the mean power spectra measured from 25  simple simulations.]{Comparison of the predicted pseudo-power spectra (dotted red curves) with the mean power spectra measured from 25  simple simulations (points with error bars; see text of Section~\ref{sec:results} for details). The upper panel shows  $\widetilde{C}_\ell^{EE}$ and the lower panel shows $\widetilde{C}_\ell^{BB}$. The predicted $B$ mode pseudo power spectrum in the case of zero input $E$ modes is also shown in the lower panel as the dotted blue curve.}
    \label{fig:theoryvsim}
\end{figure}

Fig.~\ref{fig:theoryvsim} demonstrates excellent agreement between the predicted signal and the mean pseudo-power spectra measured from the  simple signal-only map-based simulations. Also shown in Fig.~\ref{fig:theoryvsim} is the predicted pseudo-$C_\ell$ $B$-mode power spectrum for the case of zero input $E$-modes ($C^{EE}_\ell = 0$). We will compare the errors in our reconstructions of the pseudo-$C_\ell$ $B$-mode power spectra against this latter ``$B$-mode only'' prediction in order to understand the precision with which our simulated pipelines can recover the cosmological $B$-mode signal. 

%
\begin{figure}
    \centering
    \includegraphics[width=0.45\textwidth]{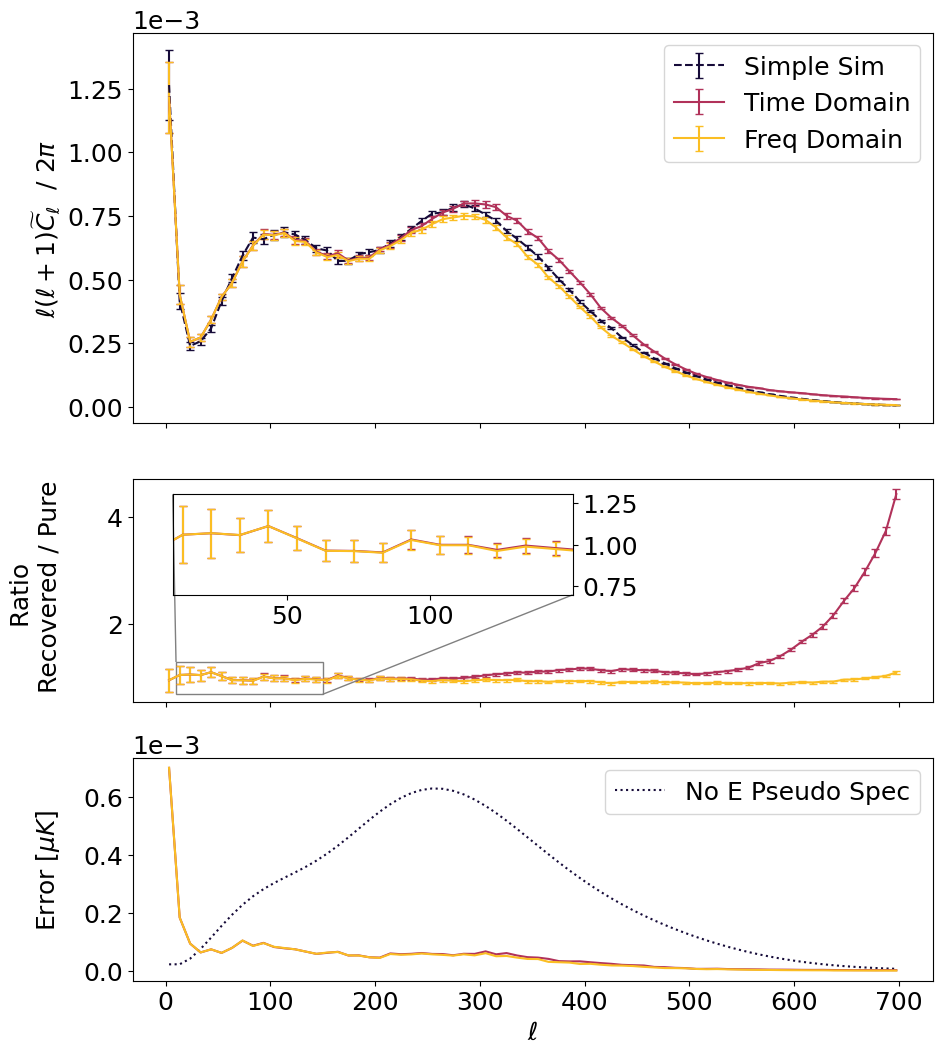}
    \caption[Recovered pseudo-$C_\ell$ $B$-mode power spectra for signal-only simulations.]{Recovered pseudo-$C_\ell$ $B$-mode power spectra for signal-only simulations. Results for the TDD (FDD) analysis are plotted as the pink (orange) points with errors. The upper panel shows the average recovered power spectra for the TDD and FDD analyses in comparison with the power spectrum measured from the simple simulations (shown as the black points with errors). The central panel shows the ratio of the recovered spectra to the  simple simulation spectra. The lower panel shows the uncertainty in the recovered spectra in comparison to the predicted $B$-mode pseudo-$C_\ell$ for the case of a null $E$-mode signal. The errors plotted in the upper and central panels show the error in the mean recovered spectra. The error plotted in the lower panel is the standard deviation amongst the 25 realisations and is indicative of the error associated with a single experiment.}
    \label{fig:noNoise_4Methods}
\end{figure}
We have run full TOD simulations for the following cases, testing the success of our four demodulation methods in each case:
\begin{itemize}
    \item Signal only.
    \item Signal + white noise.
    \item Signal + white noise + $1/f$ noise.
    \item Signal + white noise + polarised $1/f$ noise.
    \item Signal + white noise + differential gain systematic.
\end{itemize}

In all cases explored, we have found the results of differencing followed by demodulation to be indistinguishable from an analysis where we demodulate the two detectors separately and subsequently average the two sets of demodulated TODs and/or maps. This general finding holds for both the TDD and FDD approaches. For the remainder of this section, we will therefore present results for the differenced analyses only, with the understanding that the results for the corresponding single-detector analysis followed by averaging are identical. We will return to the question of whether single-detector analyses offer potential benefits compared to analyses based on differencing in Section~\ref{sec:diff_effect}.

\subsection{Signal-Only Simulations}
\label{sec:signOnly}
The results of the signal-only case are presented in Fig.~\ref{fig:noNoise_4Methods}, for both the FDD and TDD analyses. We observe excellent agreement between the $B$-mode power spectrum measured from the  simple simulations and that recovered from the FDD analysis. This level of agreement is observed over the full range of angular scales investigated. The TDD approach, on the other hand, performs comparatively poorly above multipoles, $\ell \gtrsim 300$, with an additional spurious signal seen on scales corresponding to the $\ell = 350$ acoustic peak in the $E$-mode spectrum (see Fig.~\ref{fig:theoryvsim}). We have confirmed, by running additional simulations with the input $E$-mode signal set to zero, that the spurious signal seen in the TDD recovery is due to an additional unaccounted-for $E \rightarrow B$ leakage effect inherent in that technique. 
We have also confirmed that increasing the demodulation time length causes greater leakage at high $\ell$, though we cannot ascertain the cause for this. As stated in section \ref{sec:timedemod}, the assumption of constancy over the demodulation time period should only affect recovery at very high $\ell$ values. We note that any small remaining residuals between the model and the data would possibly require further modelling and correction for high precision analyses. As this applies equally to both methods, we do not investigate corrections to this order of precision here.


\subsection{White Noise}
\label{sec:white_noise}
\begin{figure}
    \includegraphics[width=0.45\textwidth]{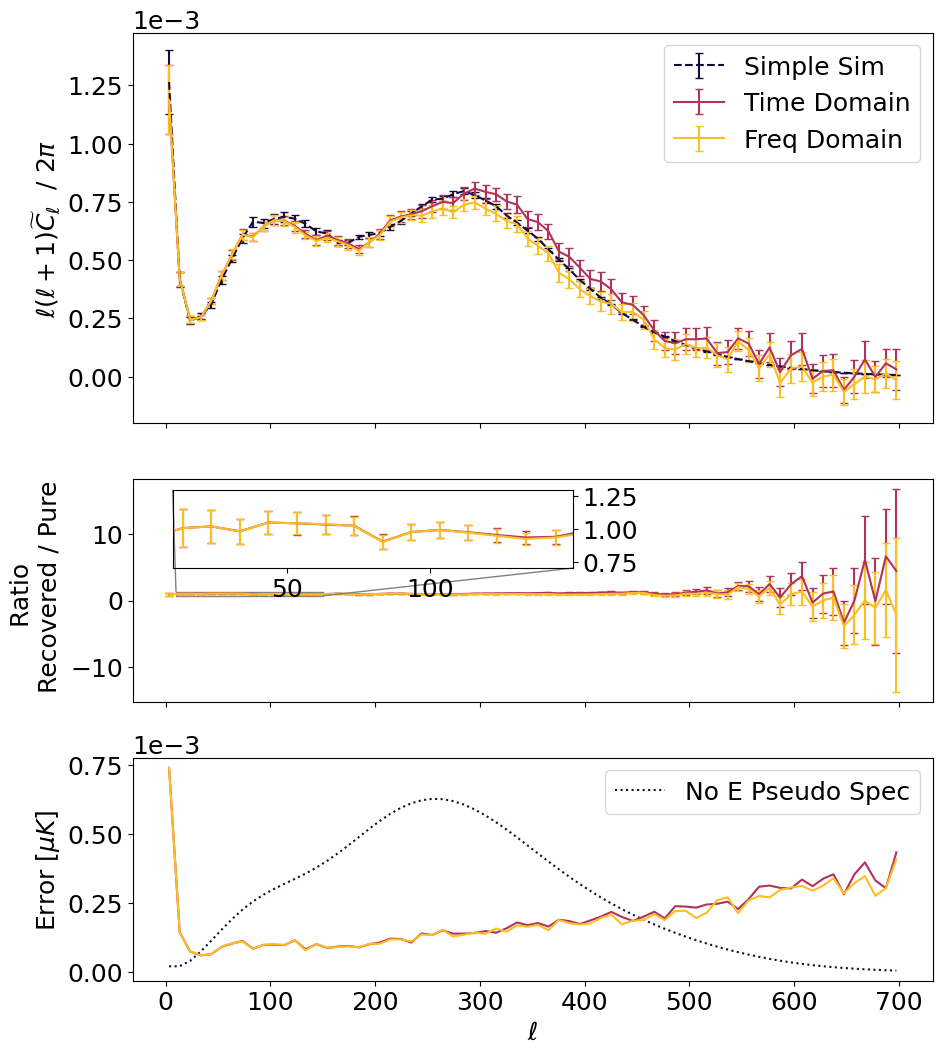}
    \caption[Recovered pseudo-$C_\ell$ $B$-mode power spectra for simulations containing white noise.]{Recovered pseudo-$C_\ell$ $B$-mode power spectra for simulations containing white noise. The upper panel shows the recovered spectra, the central panel shows the ratio of the recovered spectra to that recovered from the simple simulations and the lower panel shows the standard error compared to the predicted $B$-mode-only signal. See the caption of Fig.~\ref{fig:noNoise_4Methods} for further details of the curves plotted.}
    \label{fig:whiteNoise_4Methods}
\end{figure}

The results from the simulations containing signal and white noise components are presented in Fig.~\ref{fig:whiteNoise_4Methods}. The results are broadly consistent with what was found in the signal-only case (Fig.~\ref{fig:noNoise_4Methods}). In particular, we see the same failure of the TDD analysis for multipoles, $\ell \gtrsim 300$. The lower panel of Fig.~\ref{fig:whiteNoise_4Methods} shows that, in addition to resulting in a less accurate reconstruction of the signal, the TDD approach also results in a larger error than that seen in the FDD case. Nevertheless, we note that, for scales relevant for primordial $B$-mode measurements ($\ell \lesssim 100$) the performance is very similar between the two approaches.  
%

\subsection{$1/f$ Noise}
\begin{figure}
    \includegraphics[width=0.45\textwidth]{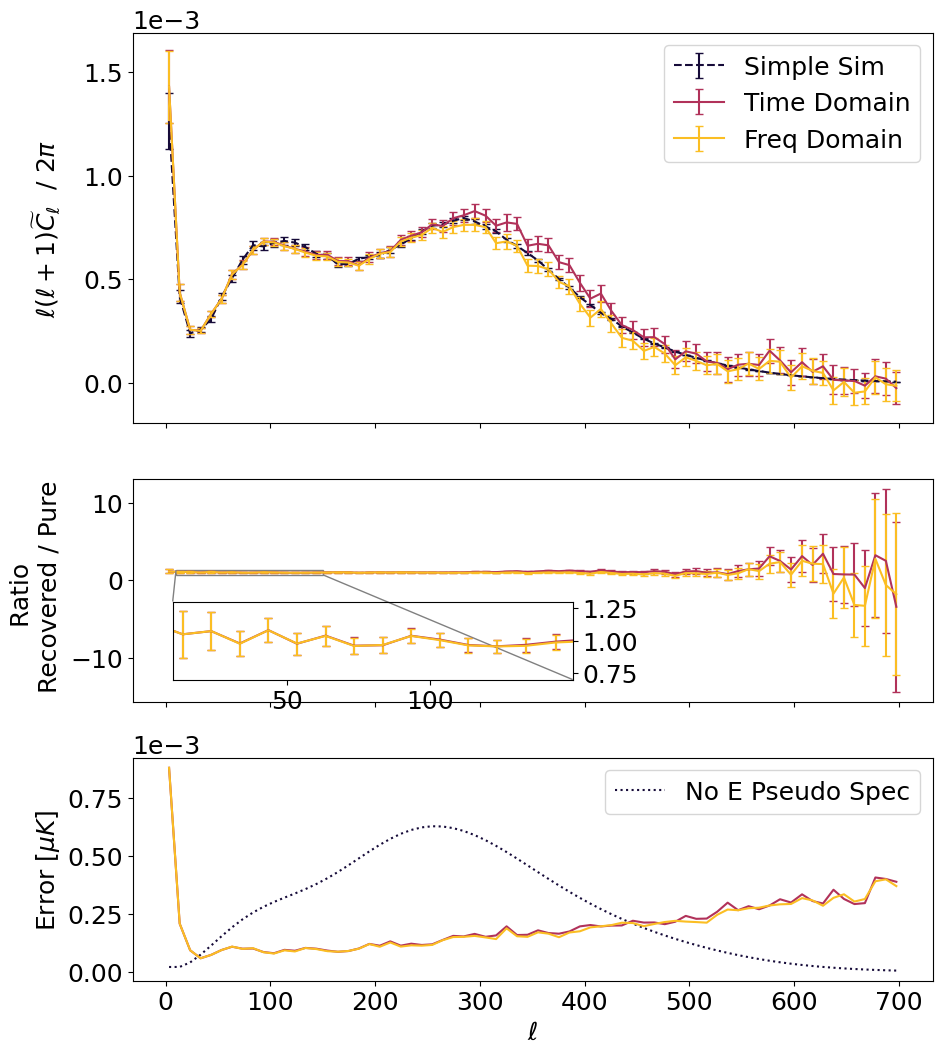}
    \caption[Recovered pseudo-$C_\ell$ $B$-mode power spectra for simulations containing both white noise and common-mode $1/f$ noise.]{Recovered pseudo-$C_\ell$ $B$-mode power spectra for simulations containing both white noise and common-mode $1/f$ noise. The upper panel shows the recovered spectra, the central panel shows the ratio of the recovered spectra to that recovered from the simple simulations and the lower panel shows the standard error compared to the predicted $B$-mode-only signal. See the caption of Fig.~\ref{fig:noNoise_4Methods} for further details of the curves plotted.}
    \label{fig:white-1f-Noise_4Methods}
\end{figure}

The results of the simulations including both white noise and common-mode $1/f$ atmospheric noise are presented in Fig.~\ref{fig:white-1f-Noise_4Methods}. Qualitatively, we see no difference between these and the results from the white-noise only case, demonstrating the well-known utility of a CRHWP in mitigating the effects of common-mode $1/f$ noise. We also see, from the agreement between the TDD and FDD results on large angular scales, that the success with which a CRHWP mitigates $1/f$ noise (which predominantly affects large scales, low multipoles) is independent of the demodulation approach employed.   

\label{1overfnoise}

\subsection{Polarised $1/f$ Noise}
\begin{figure}
    \includegraphics[width=0.45\textwidth]{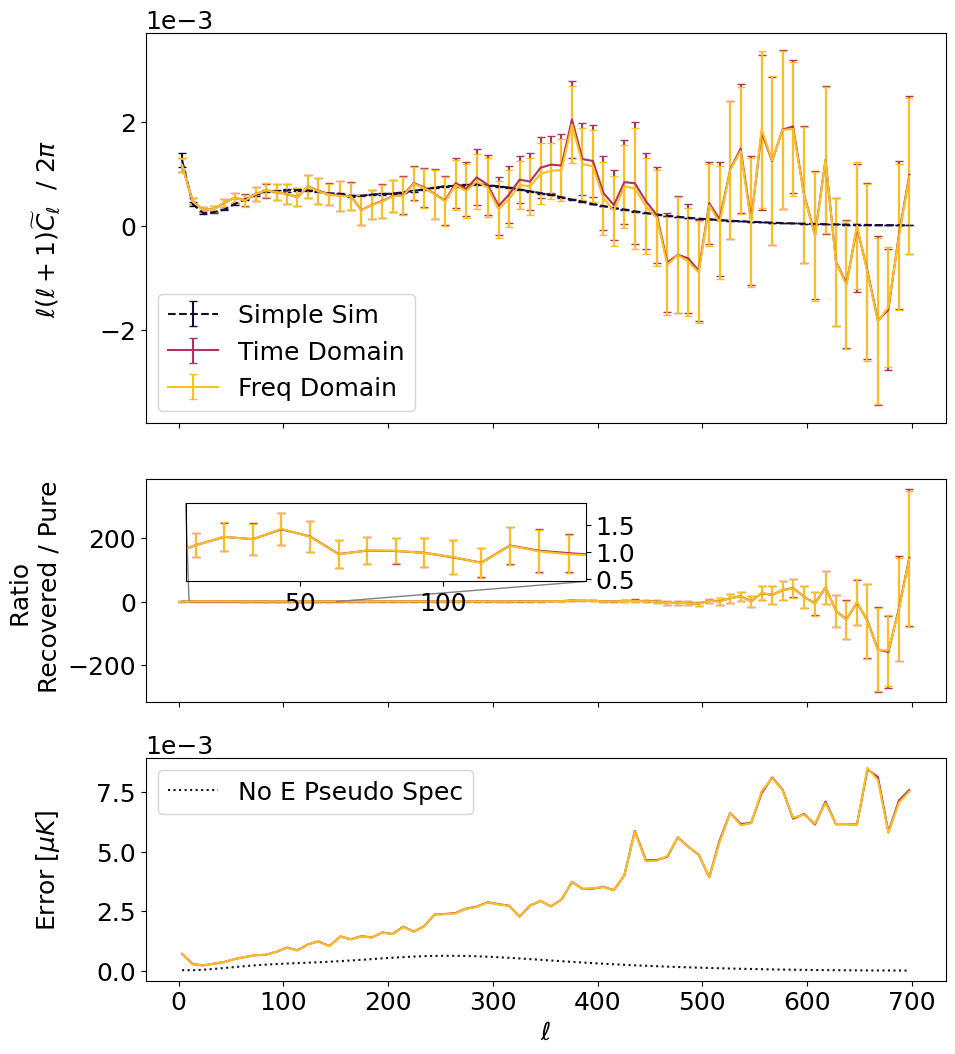}
    \caption[Recovered pseudo-$C_\ell$ $B$-mode power spectra for simulations containing white noise and polarised $1/f$ noise, with an assumed polarisation fraction of $f_{\rm pol} = 0.5$.]{Recovered pseudo-$C_\ell$ $B$-mode power spectra for simulations containing white noise and polarised $1/f$ noise, with an assumed polarisation fraction of $f_{\rm pol} = 0.5$. The common mode $1/f$ contribution to the noise has been turned off to highlight the effect of the polarised noise. The upper panel shows the recovered spectra, the central panel shows the ratio of the recovered spectra to that recovered from the  simple simulations and the lower panel shows the standard error compared to the predicted $B$-mode-only signal. See the caption of Fig.~\ref{fig:noNoise_4Methods} for further details of the curves plotted.}
    \label{fig:white-polnoise_4Methods}
\end{figure}

As an example of a systematic effect/additional noise source that is not mitigated by the CRHWP, we have also considered the case of a polarised $1/f$ component in the atmosphere. We note that though HWPs do not mitigate for this, other techniques (such as more sophisticated map making techniques) can do so. This paper focuses on the utility of a CRHWP, and so such methods of mitigation are beyond the scope of this work. The power spectra recovered from these simulations are shown in Fig.~\ref{fig:white-polnoise_4Methods}, again for both the TDD and FDD methods. Due to the cross-correlation approach to removing noise (see Section~\ref{sec:analysis}), the recovered power spectra remain unbiased. However, it can be seen, from the bottom panel of Fig.~\ref{fig:white-polnoise_4Methods}, that the precision of the recovery is severely degraded by the polarised $1/f$ component in all cases. Again we see no discernible difference between the performance of the TDD and FDD pipelines for this case. While the additional $E \rightarrow B$ leakage on intermediate scales ($\ell \sim 350$) associated with the TDD approach (see Fig.~\ref{fig:noNoise_4Methods} and Section~\ref{sec:signOnly}) appears less prominent in Fig.~\ref{fig:white-polnoise_4Methods}, we stress that this is solely the result of the large increase in errors (due to the polarised $1/f$ noise) dominating over this leakage effect, i.e. the absolute level of leakage is similar to that seen in the previous simulations (Figs.~\ref{fig:noNoise_4Methods} -- \ref{fig:white-1f-Noise_4Methods}). 

\subsection{Differential Gain}
\begin{figure}
    \includegraphics[width=0.45\textwidth]{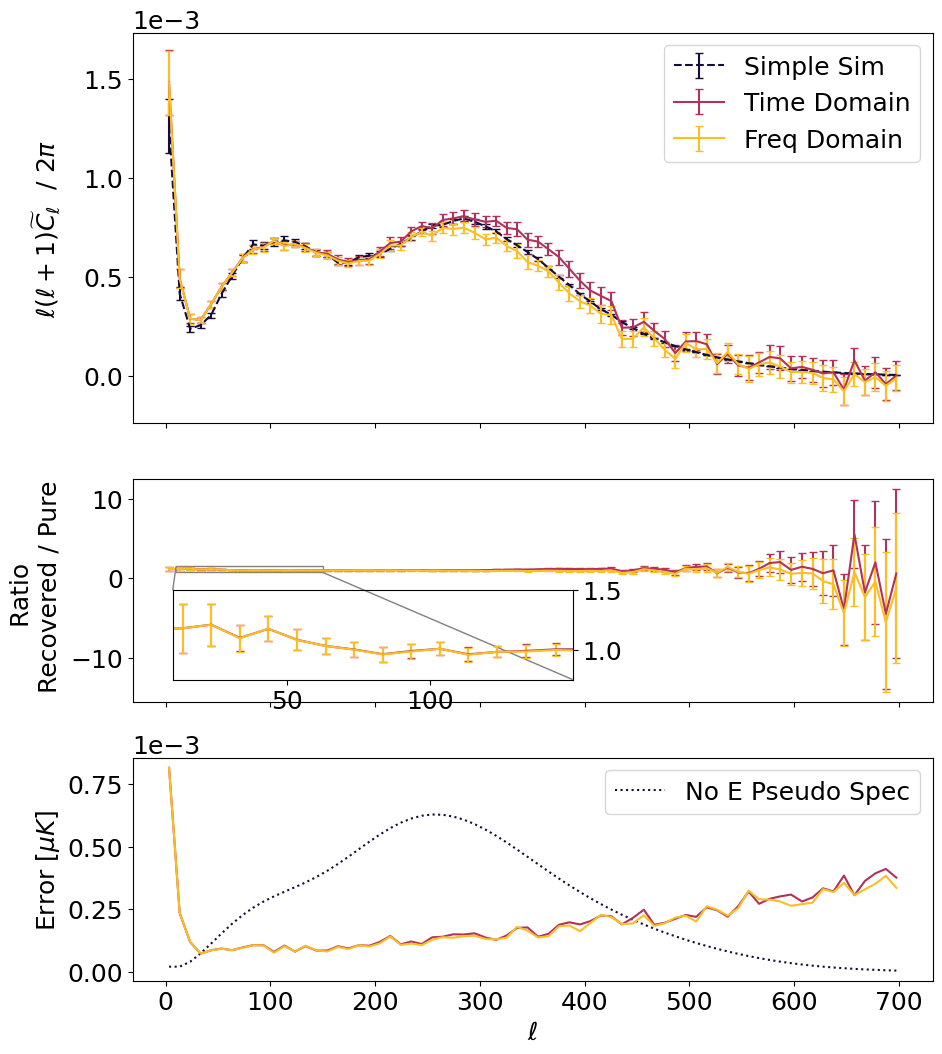}
    \caption[Recovered pseudo-$C_\ell$ $B$-mode power spectra for simulations containing both white noise and a 1\% differential gain error between the two detectors.]{Recovered pseudo-$C_\ell$ $B$-mode power spectra for simulations containing both white noise and a 1\% differential gain error between the two detectors. The upper panel shows the recovered spectra, the central panel shows the ratio of the recovered spectra to that recovered from the  simple simulations and the lower panel shows the standard error compared to the predicted $B$-mode-only signal. See the caption of Fig.~\ref{fig:noNoise_4Methods} for further details of the curves plotted.}
    \label{fig:white-gain-Noise_4Methods}
\end{figure}

The results of the simulations including both white noise and a differential gain error are presented in Fig.~\ref{fig:white-gain-Noise_4Methods}. Comparing with the results for the simulations containing only signal and white noise (Fig.~\ref{fig:whiteNoise_4Methods}), we see the expected result that the presence of the CRHWP effectively mitigates the $T \rightarrow P$ leakage associated with differential gain errors (in the absence of other systematic effects).  
\label{diff_gain}

As with all of the results presented in this section, we have observed no differences between analyses based on detector differencing and analyses based on single-detector demodulation, where the latter is followed by averaging of the timestreams/maps from the two detectors. For the TDD case, this is partly due to the fact that, subsequent to detector differencing, we retain the $T$ row of equation~(\ref{eq:tdd}) for the demodulation/map-making step despite the fact that the temperature signal should have nominally already been rejected by the differencing. This approach mirrors the analysis of the POLARBEAR data presented in \cite{2014ApJ...794..171P} and its proposed extensions aimed at rejecting systematics with specific spin properties, as suggested in \cite{mccallum21}. We have also confirmed that the $B$-mode recovery for the differenced TDD analysis is degraded, in the presence of the gain systematic, if the demodulation/map-making step is restricted to recover only the $Q$ and $U$ Stokes parameters. 

\section{Detector differencing vs single-detector analysis}
\label{sec:diff_effect}
In all of the cases presented in the previous section, we have found the results of differencing followed by demodulation to be indistinguishable from an analysis where the two detectors are demodulated separately and the two sets of demodulated TODs are then averaged. This general finding remains true for both the TDD and FDD approaches. However there is a difference in how information from multiple detectors is combined in these two approaches -- in the differencing analysis, information from multiple detectors is combined at the instantaneous timestream level whereas, for analyses based on demodulating individual detectors, the combination of information from multiple detectors can be postponed until a later stage of the analysis -- the most obvious example of which would be the subsequent co-addition of single-detector $T$, $Q$ and $U$ maps. This ability to measure the $T$, $Q$ and $U$ Stokes parameters from single detectors in isolation is often presented as a key advantage of fast modulation schemes \cite[e.g.][]{maxipol_hwp, Essinger_Hileman_2016}.

\begin{figure}
    \includegraphics[width=0.45\textwidth]{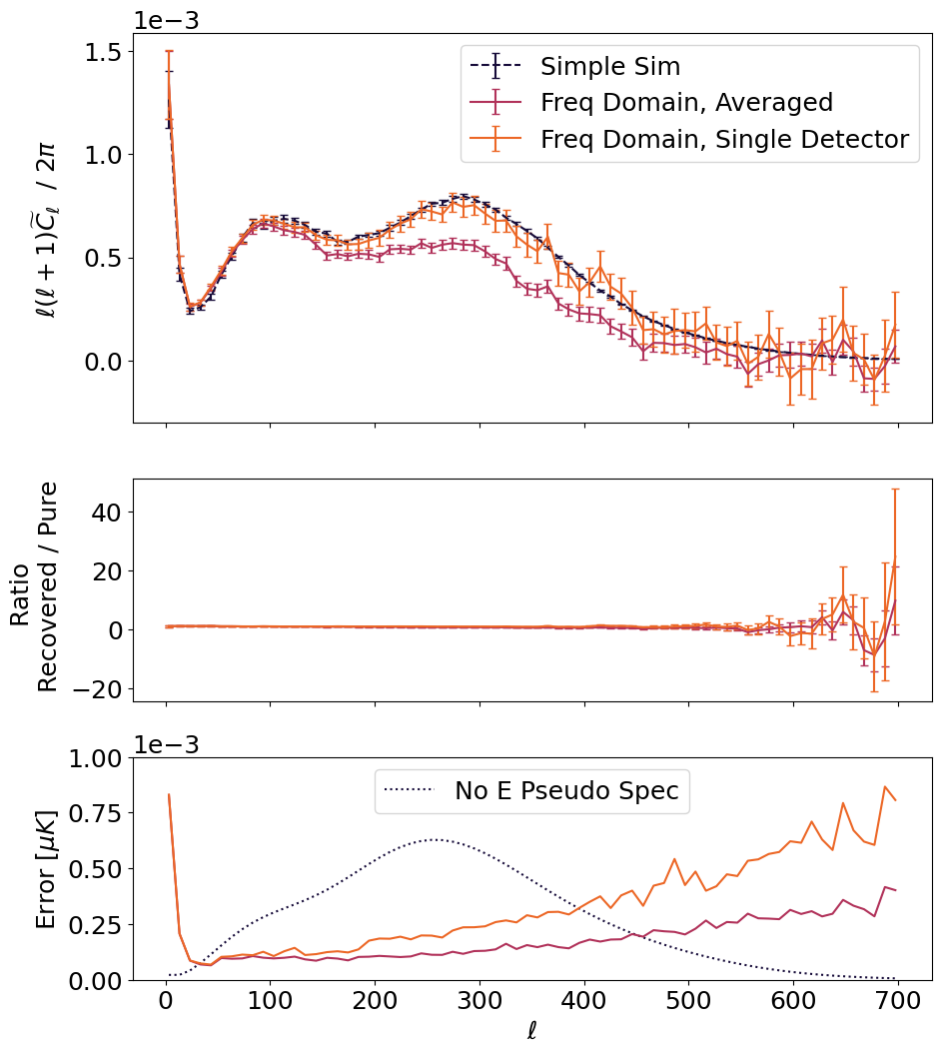}
    \caption[Recovered pseudo-$C_\ell$ $B$-mode power spectra for simulations containing both white noise and a pointing error of 1.7 arcminutes.]{Recovered pseudo-$C_\ell$ $B$-mode power spectra for simulations containing both white noise and a pointing error of 1.7 arcminutes, in both latitude and longitude directions, between the two detectors. The purple and orange points with errors show the results from the FDD averaged and FDD single detector analyses respectively. The single detector analysis refers to the case where the demodulated TOD from only one of the two detectors has been retained. The upper panel shows the recovered spectra, the central panel shows the ratio of the recovered spectra to that recovered from the simple simulations and the lower panel shows the standard error compared to the predicted $B$-mode-only signal. See the caption of Fig.~\ref{fig:noNoise_4Methods} for further details of the curves plotted.}
    \label{fig:white-pointing-Noise_freqOnly}
\end{figure}

We can use the example of a differential pointing systematic (an instrumental systematic that is not mitigated by a CRHWP) to demonstrate how single-detector data processing could, in principle, be beneficial. In Fig.~\ref{fig:white-pointing-Noise_freqOnly} we present the results of simulations including both white noise and a differential pointing error of 1.7 arcminutes, in both latitude and longitude directions, added to each of the two detectors. We identify a target area of the sky, and then add a 1.7 arcminute offset in both RA and declination to both detectors, in opposite directions.

For clarity, we note that, although it is usually $T \rightarrow P$ leakage that is considered when examining differential pointing effects, differential pointing also causes $P \rightarrow P$ leakage~\citep[see e.g.][]{mccallum21}. In the case considered here, the $T \rightarrow P$ leakage is removed by the modulation of the CRHWP, and the remaining bias seen in Fig.~\ref{fig:white-pointing-Noise_freqOnly} is due to $P \rightarrow P$ leakage. We have verified numerically using our simulations that the leakage we are seeing is not $T \rightarrow P$ leakage.

The FDD approach is presented here, as the more successful of the two. Rather than presenting a differenced case however, the averaged case is contrasted against a 'single detector' analysis.  
For the single-detector demodulation case, we do not subsequently average the demodulated TODs. Rather, we simply discard one of the TODs and proceed with the subsequent map-making and power spectrum estimation using only the data from a single detector. As expected, the analyses that use only the data from a single detector do not suffer from the effects of the differential pointing error. They do, of course, suffer from an increased random error in the recovery due to the fact that 50\% of the data has been discarded (Fig,~\ref{fig:white-pointing-Noise_freqOnly}, lower panel). 

Note that, in this case, it is not the modulation of the incoming polarisation signal that is beneficial here (fast modulation does not mitigate the effects of a differential pointing error). Rather it is the fact that, by enabling observations of the same part of the sky with a range of polarisation sensitivity directions, the CRHWP allows for the incoming signal detected by a single (linearly polarised) detector to be separated into its constituent Stokes parameter components. This particular capacity (to measure all three Stokes parameter from a single detector) is not specific to a continuously rotating HWP. It can also be achieved using a stepped HWP or indeed, in the absence of a HWP, using sufficiently frequent stepped rotations of the telescope/receiver boresight angle. 

The example of differential pointing presented here is representative of a wider class of systematic effects that can cause $T \rightarrow P$ and/or $P \rightarrow P$ leakage and which are not mitigated by a CRHWP \citep[see e.g.][]{hu03, odea07, shimon08, HWPOverheat_Michael, mccallum21}. While recovering Stokes parameters from individual detectors can potentially help, it does not completely solve the problems associated with such systematic effects. In particular one is still left with the issue of how to reject mismatches in the responses of different detectors when combining information from multiple detectors, albeit at a later stage in the analysis (e.g. during the co-addition of single-detector maps). 

With the large number of detectors being deployed on upcoming surveys, there may be additional computational implications to be considered. For example, if detectors are differenced when creating maps, then there are fewer maps to store and co-add. If the summed timestreams are not of interest for a particular experiment and one's calibration and instrument model are performed and constructed appropriately, one could store only the differenced timestreams. This factor of two could be significant given the large amounts of data expected from future ground based surveys, expected to be tens of petabytes (PB) for forthcoming experiments.  


\begin{figure}
    \centering
    \includegraphics[width=0.45\textwidth]{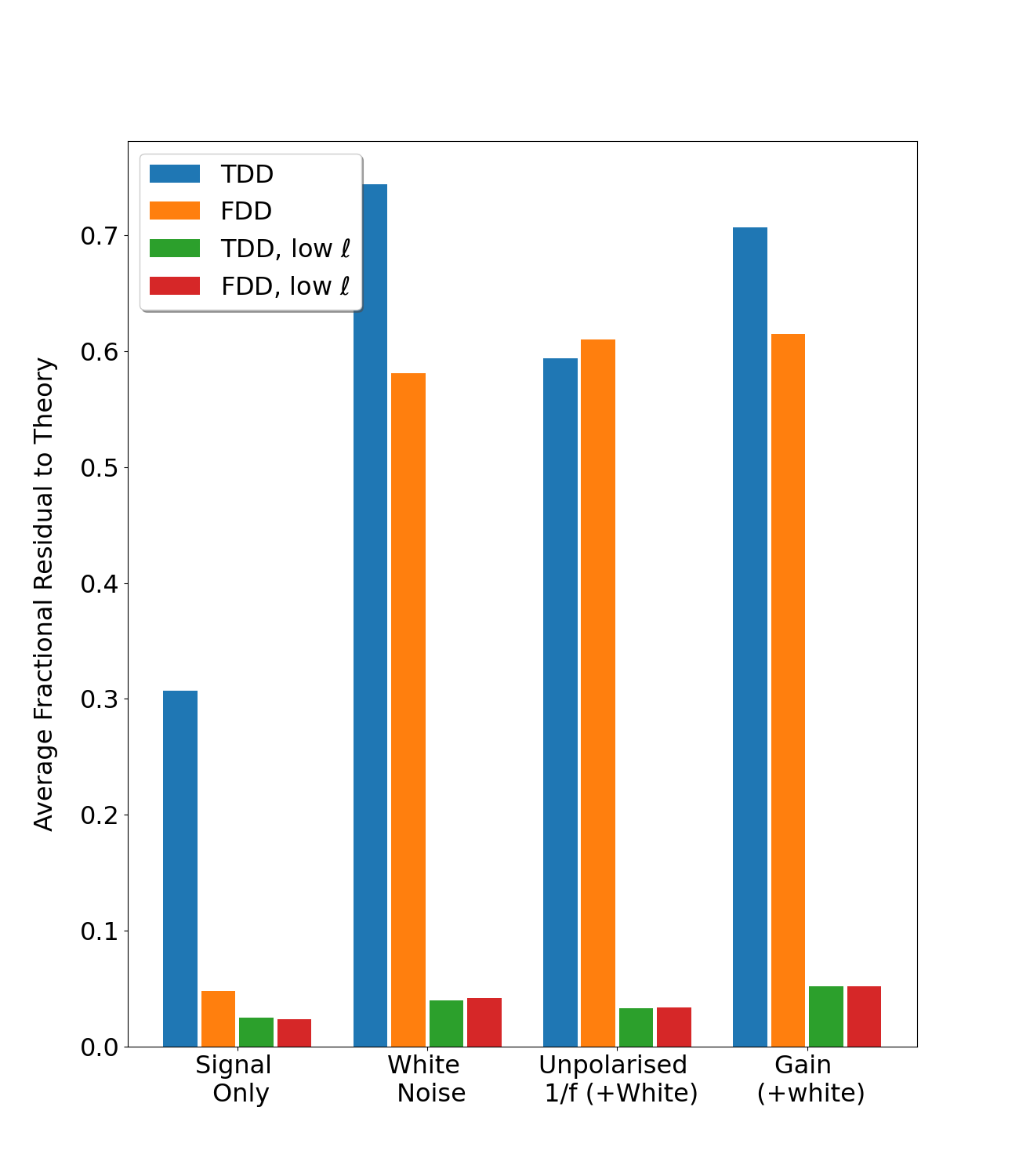}
    \caption[The success of each demodulation method, in each regime, quantified as the magnitude of the residual between the recovered signal and the predicted theoretical pseudo spectrum.]{The success of each method, in each regime, quantified as the magnitude of the residual between the recovered signal and the predicted theoretical pseudo spectrum, averaged over multipoles, $\ell$. It can be seen that, for the full multipole range ($10 < \ell < 700$; labelled ``TDD'' and ``FDD''), the TDD technique is usually slightly inferior to the FDD technique. The only exception to this is the case of unpolarised $1/f$ noise, for which the difference in performance is very slightly in favour of the TDD technique. These difference are all well within the standard error (see Fig.~\ref{fig:err_bars}), and so we maintain that the additional leakage seen in the TDD technique suggests the FDD technique is superior. However, for scales relevant for primordial $B$-mode studies ($10 < \ell < 150$; labelled ``low $\ell$''), we see a negligible difference between the performance of the two methods. The leakage effect is not present on these scales.}
    \label{fig:resi_bars}
\end{figure}

\begin{figure}
    \centering
    \includegraphics[width=0.45\textwidth]{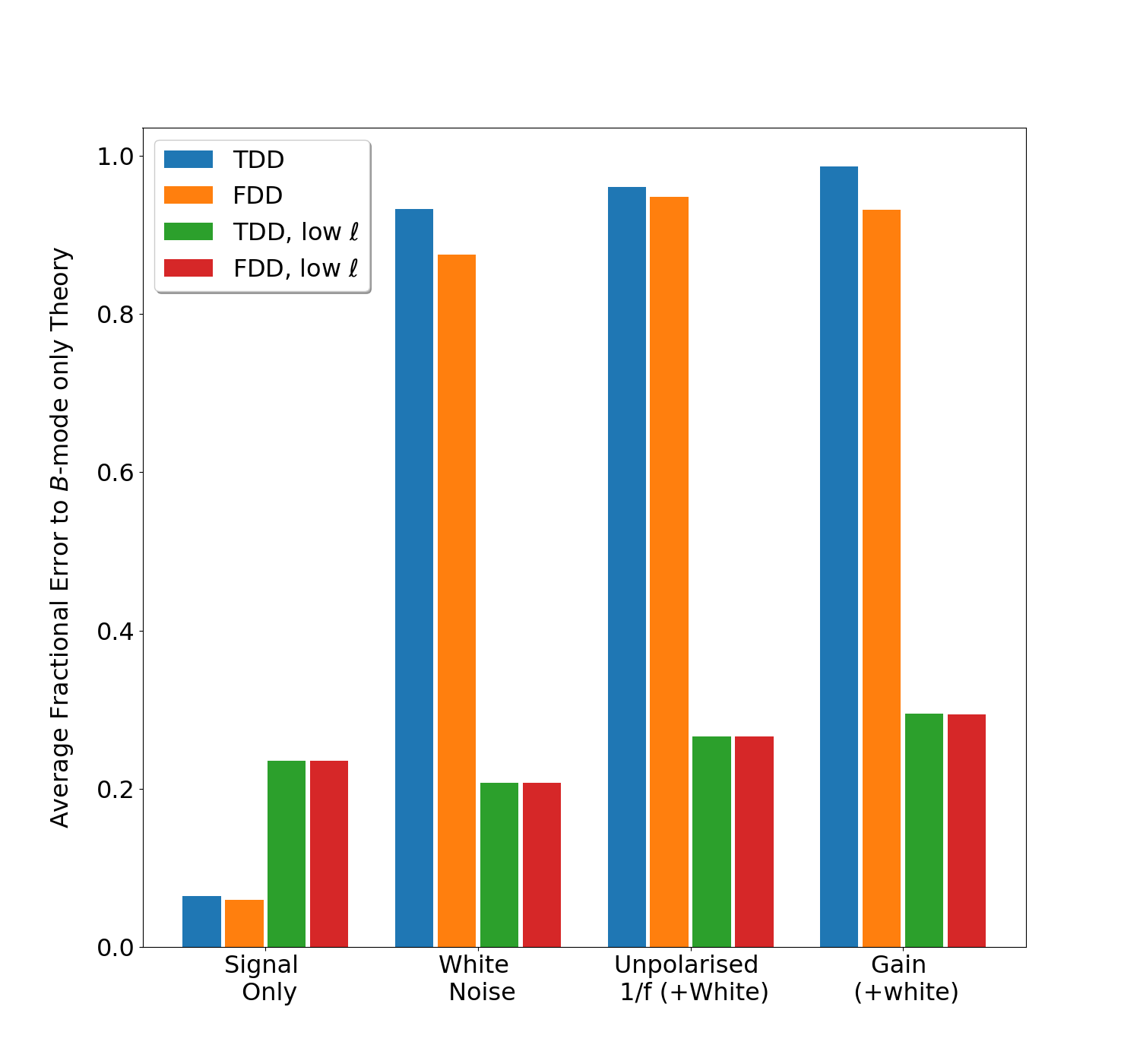}
    \caption[The error in each demodulation method, in each noise regime, calculated as the standard error over 25 runs, and averaged over multipoles.]{The error in each method, in each noise regime, calculated as the standard error over 25 runs, and averaged over multipoles, $\ell$. Errors are plotted as a fraction of the predicted (theoretical) $B$-mode pseudo-$C_\ell$ spectrum in the absence of $E$-modes (as described in section \ref{sec:noise}) in order to quantify the relative size of the uncertainty in our measurements with respect to the amplitude of the expected $B$-mode signal which we aim to recover. As in Fig.~\ref{fig:resi_bars}, the TDD technique is found to be inferior to the FDD technique over the entire multipole range. However, we see a negligible difference in the performance of the two techniques for the low $\ell$ regime.}
    \label{fig:err_bars}
\end{figure}

\section{Conclusions}
\label{sec:conclusion}
Several forthcoming CMB experiments will use CRHWPs to aid in the robust recovery of the $B$-mode polarisation signal. In this paper, we have investigated how different analysis choices affect the recovered signal when a CRHWP is present. In particular, we have compared the performance of two different approaches to demodulating the polarisation signal, and have explored the trade-offs involved in deciding if data from detector pairs co-located on the focal plane should be differenced instantaneously in the timestream or not. 

We have tested both the classical lock-in demodulation technique (Frequency domain demodulation, FDD), in addition to a real-space ``time-domain" approach (Time domain demodulation, TDD), using five different scenarios in terms of the assumed noise properties and levels of instrumental systematic effects. 

In all cases, the lock-in method performed better than the TDD method in terms of the accuracy with which the $B$-mode power spectrum was recovered over the full range of angular scales measured. The only exception to this was the polarised $1/f$ noise case, where the level of recovery was very similar between the two approaches. In this particular case, the heavily degraded precision of the recovery is driven by the polarised $1/f$ noise which is not modulated by the CRHWP. 

One persistent feature seen in the TDD approach, and which is not present in the FDD method, is an $E \rightarrow B$ leakage effect arising at intermediate multipoles $\ell\sim350$. While we have yet to identify the origin of this effect, it is highly unlikely to be related to the approximation inherent in the TDD approach that the on-sky fields are taken to be constant over the demodulation timescale, $\tau_{\rm dm}$. Given the sampling frequency and scan speed used in our simulations, any effects associated with this approximation would be expected to manifest at much higher multipoles, $\ell > 2800$. 

In terms of the recovery of the signal on the relatively large angular scales where searches for primordial $B$-modes are focused, we find that the FDD and TDD approaches perform equally well. This offers a potentially powerful consistency check for future experiments targeting this signal and employing a CRHWP.   

We summarise these conclusions on the relative performance of the FDD and TDD methods in Figs.~\ref{fig:resi_bars} and \ref{fig:err_bars}. In Fig.~\ref{fig:resi_bars} we plot the residual between the mean recovered $B$-mode pseudo-$C_\ell$ and the theoretical prediction, for several of the experimental scenarios investigated. We present results averaged over both the full range of angular scales measured ($10 < \ell < 700$), and over a restricted low-$\ell$ range appropriate for primordial $B$-modes ($10 < \ell < 150$). Similarly, Fig.~\ref{fig:err_bars} shows the random errors in the recovered spectra for the same experimental scenarios and multipole ranges. 

For both the TDD and FDD approaches, we have also compared the results of analyses that include a prior detector-differencing step with analyses that recover all three Stokes parameters ($T$, $Q$, $U$) from individual detectors in isolation. For most of the scenarios investigated, we find that this choice makes no difference to either the accuracy or precision of the recovered $B$-mode power spectra. However, for instrumental systematic effects that are not mitigated by the CRHWP (such as differential pointing), we have argued that measuring all three Stokes parameters from individual detectors could provide a potential advantage by way of moving the step of detector combination to a later point in the analysis pipeline. The extent to which this is beneficial is currently not clear and should be investigated further. At the same time, we also note that such an approach places a greater demand on storage requirements for future experiments, which are already significant.

We expect our results to be useful for forthcoming experiments planning on using CRHWP including Simons Observatory \citep{SO_Main} and LiteBIRD \citep{LiteBIRD_1}. There is significant scope for future work to build upon our conclusions by incorporating additional systematics into the simulation pipelines described here, in particular, non-idealities in the HWP such as differential transmittance and frequency-dependent effects \citep[e.g.][]{2021MNRAS.502.4526D, Giardiello22}. 


\section*{Acknowledgements}
MR, MLB and DBT acknowledge support from the Science and Technology Facilities Council (STFC grant numbers ST/T506291/1, ST/P000592/1, ST/X006336/1 and ST/X006344/1).


\section*{Data Availability}

Data is available from the authors on request. 



\bibliographystyle{mnras}
\bibliography{paper} 




\appendix

\section{Effect of demodulation time-step in the TDD method.}

\begin{table*}
\centering
\begin{tabular}{l|l|ll|ll|}
\cline{2-6}
 & \multirow{2}{*}{\textbf{\begin{tabular}[c]{@{}l@{}}Demod Time \\ Step\end{tabular}}} & \multicolumn{2}{l|}{\textbf{TDD}}                           & \multicolumn{2}{l|}{\textbf{FDD}}                           \\ \cline{3-6} 
 &                                                                                      & \multicolumn{1}{l|}{\textit{$10 < \ell < 150$}} & $10 < \ell < 700$ & \multicolumn{1}{l|}{\textit{$10 < \ell < 150$}} & $10 < \ell < 700$ \\ \cline{2-6} 
 & \textbf{$3 / f_{\rm{samp}} $}                                                             & \multicolumn{1}{l|}{0.0347}                 & 0.8399        & \multicolumn{1}{l|}{0.0293}                 & 0.0478        \\ \cline{2-6} 
 & \textbf{$5 / f_{\rm{samp}} $}                                                             & \multicolumn{1}{l|}{0.0360}                 & 0.3099        & \multicolumn{1}{l|}{0.0367}                 & 0.0495        \\ \cline{2-6} 
 & \textbf{$7 / f_{\rm{samp}} $}                                                             & \multicolumn{1}{l|}{0.0377}                 & 1.2020        & \multicolumn{1}{l|}{0.0385}                 & 0.0508        \\ \cline{2-6} 
 & \textbf{$9 / f_{\rm{samp}} $}                                                             & \multicolumn{1}{l|}{0.0308}                 & 1.9925        & \multicolumn{1}{l|}{0.0305}                 & 0.0492        \\ \cline{2-6} 
\end{tabular}
\caption{Table to show the effect of varying the demodulation time length on each of the methods, across two ranges of $\ell$. The values in this table are average fractional residuals; the residual at each $\ell$ value is calculated as described in equation \ref{eq:demodFracRes}, and the average over a given $\ell$ range is then taken. The FDD method does not utilise the demodulation time length parameter, and is included here as a point of reference.}
\label{tab:varydemodstep}
\end{table*}

This appendix outlines our investigation into the origin of the $E \rightarrow B$ leakage effect in the TDD method. Whilst we do not at this time fully understand the origin of the effect, we have nevertheless been able to identify a set of parameter choices for the TDD analysis that minimises it. We also note that the underlying assumption made in the TDD case should not be the origin of this effect on these scales, as per the discussion at the end of section 2.1. 

Of all the parameters in our simulations, we found that the demodulation time length had the greatest impact on the $E \rightarrow B$ leakage effect. Increasing the value of $\tau_{dm}$ had a noticeable effect on the average fractional residual. The fractional residual at each value of $\ell$ is calculated as follows:

\begin{equation}
    \delta C_{\ell} = \frac{C_{\ell}^{demod} -  C_{\ell}^{theory}}{C_{\ell}^{theory}}
    \label{eq:demodFracRes}
\end{equation}

where $C_{\ell}^{demod}$ is the demodulated power spectrum and $C_{\ell}^{theory}$ is the theoretical pseudo-spectrum, calculated as described in section \ref{sec:results}. The average over a given $\ell$ range is then taken to produce the values in table \ref{tab:varydemodstep}. We have included the FDD columns as a control measure – the numbers in these columns indicate the expected scatter due to realisation dependence, given the FDD method does not utilise $\tau_{dm}$. 

From the $10 < \ell < 700$ TDD column in table \ref{tab:varydemodstep}, we see that the choice of $\tau_{dm} = 5/f_{\rm{samp}}$ minimises the impact of the leakage effect. Thus this value is adopted throughout the paper. At the lower $\ell$ range, the leakage effect is insignificant. 

\begin{figure}
    \centering
    \includegraphics[width=0.45\textwidth]{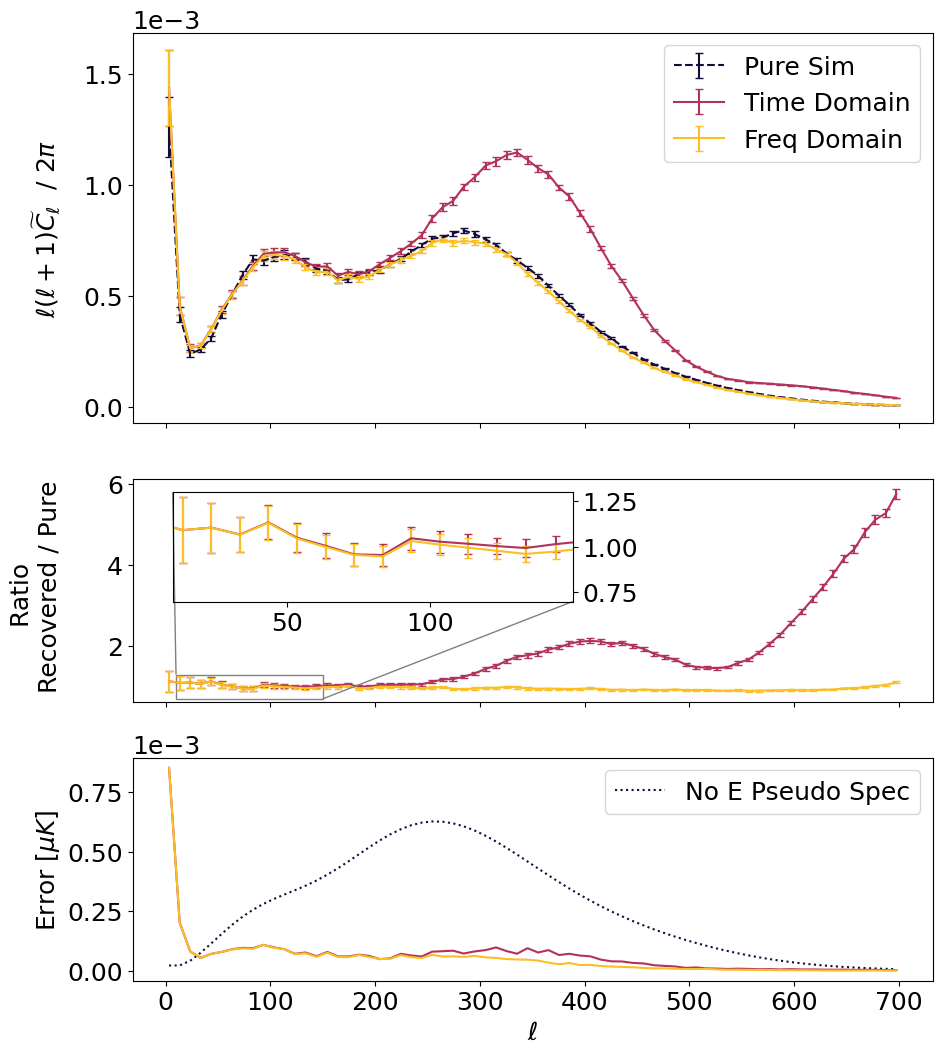}
    \caption{TDD with a demodulation time length of $3/f_{samp}$. This seems to be too few points to a time-step, with the leakage very noticeable.}
    \label{fig:demod3}
\end{figure}

\begin{figure}
    \centering
    \includegraphics[width=0.45\textwidth]{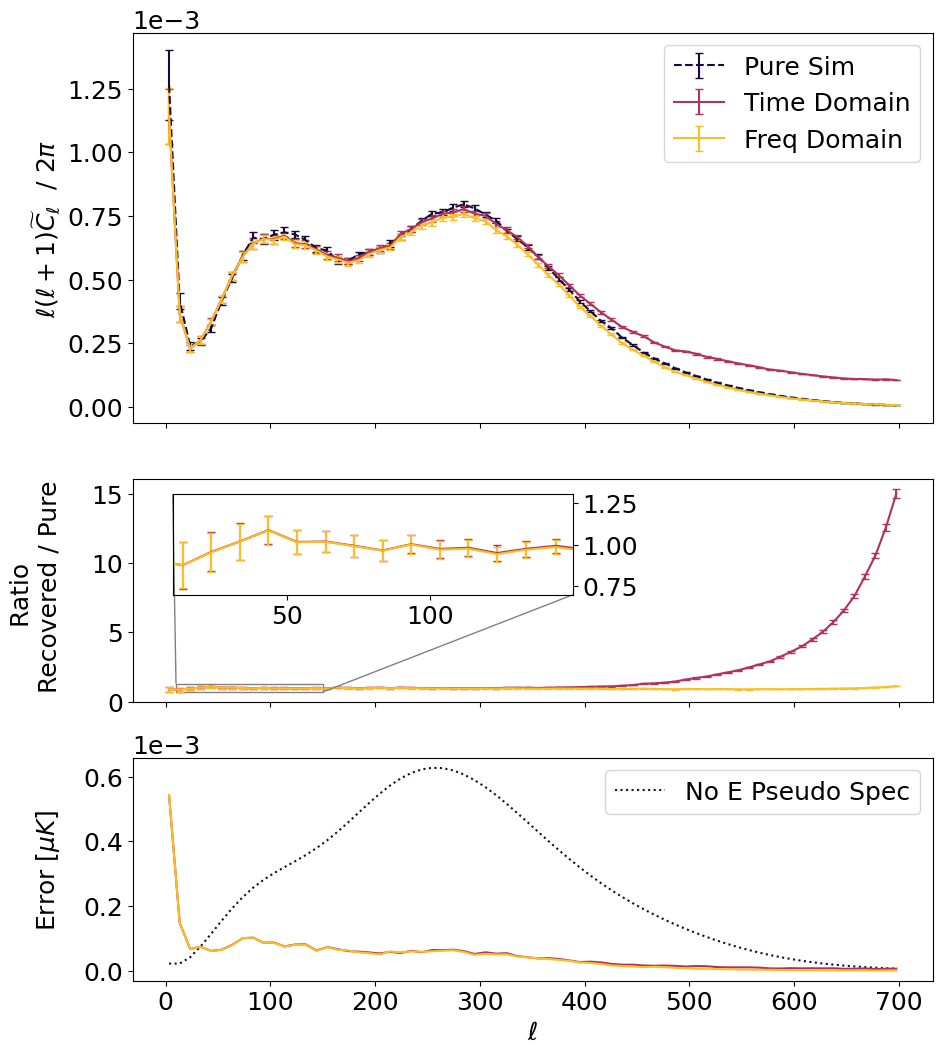}
    \caption{TDD with a demodulation time length of $7/f_{samp}$}
    \label{fig:demod7}
\end{figure}

\begin{figure}
    \centering
    \includegraphics[width=0.45\textwidth]{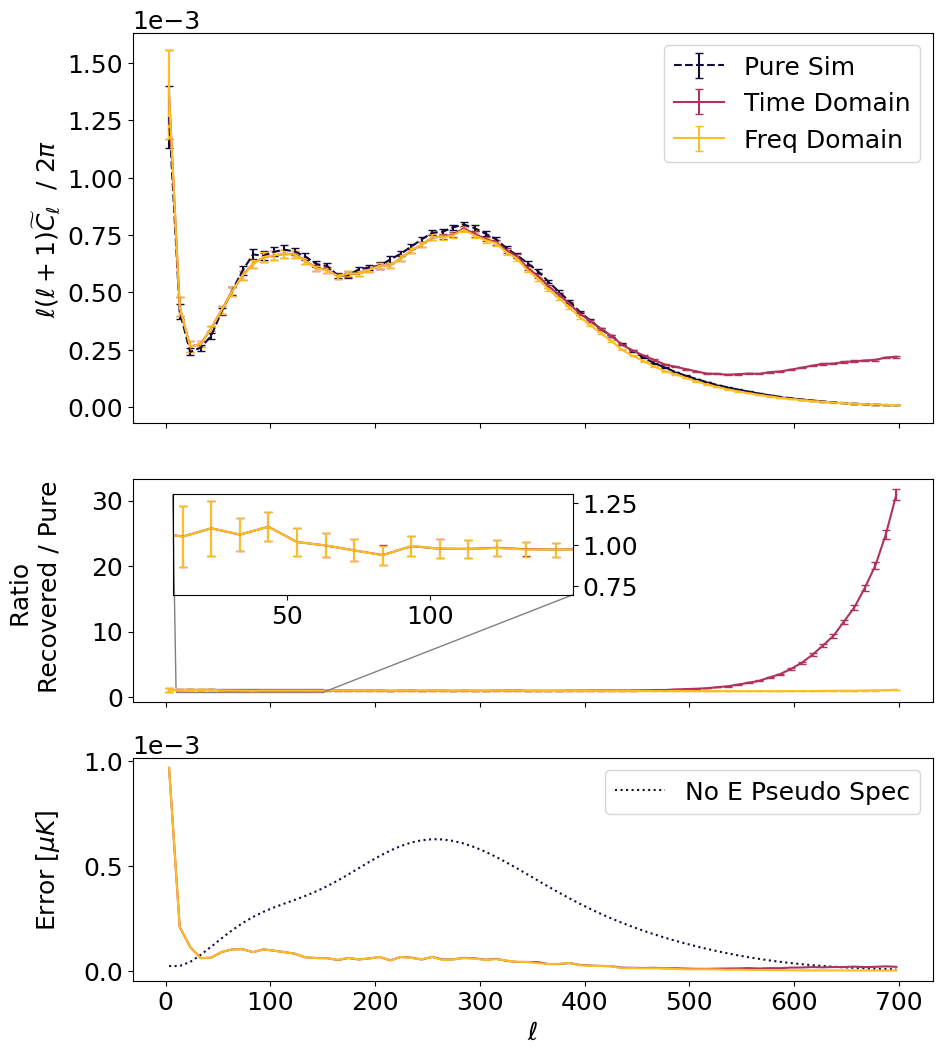}
    \caption{TDD with a demodulation time length of $9/f_{samp}$}
    \label{fig:demod9}
\end{figure}

Figures \ref{fig:demod3}, \ref{fig:demod7} and \ref{fig:demod9} show the varying demodulation time steps. It can be seen that increasing this value further decreases the success of recovery at higher-$\ell$, and that too low a value results in large amounts of leakage.

\bsp	
\label{lastpage}
\end{document}